\documentclass[twocolumn,showpacs]{revtex4}

\def\nh4{$\alpha$-(BEDT-TTF)$_2$NH$_4$Hg(SCN)$_4$}

\usepackage{graphicx}
\usepackage{dcolumn}
\usepackage{amsmath}
\begin{document}

%@twocolumnfalse\endcsname
\title{Charge-density Waves Survive the Pauli Paramagnetic Limit}
\author{R.D.~McDonald$^1$, N.~Harrison$^1$, L.~Balicas$^2$, 
K.~H.~Kim$^1$,
J.~Singleton$^1$ and X.~Chi$^1$
}
\address{$^1$
National High Magnetic Field Laboratory, LANL, MS-E536, Los
Alamos, New Mexico 87545\\$^2$
National High Magnetic Field Laboratory, Florida State University, 1800 E. Paul
Dirac Drive, Tallahassee, Florida 32310
}
%\date{\today}
%\maketitle

\begin{abstract}
Measurements of the resistance of single crystals of 
(Per)$_2$Au(mnt)$_2$ have been made at 
magnetic fields $B$ of up to 45~T, exceeding the
Pauli paramagnetic limit of $B_{\rm P}\approx 37$~T. 
The continued presence of
non-linear charge-density wave electrodynamics at $B \geq 37$~T 
unambiguously establishes the survival of the charge-density wave
state above the Pauli paramagnetic limit, and the likely emergence of
an inhomogeneous phase analogous to that anticipated to occur in
superconductors.
\end{abstract}

\pacs{71.45.Lr, 71.20.Ps, 71.18.+y}
%]\narrowtext
\maketitle 

Fundamental changes can occur within paired-electron condensates
subjected to intense magnetic fields~\cite{malo1,chaikin1,singleton1}. 
If the state is a spin-singlet (with electron spins opposed), as in
charge-density waves (CDWs)~\cite{gruner1} and {\it s}- and 
{\it d}-wave superconductors~\cite{tinkham1}, 
the energy of the partially
spin-polarized electrons of the uncondensed metal eventually becomes
lower than condensate energy above a characteristic field known as the
Pauli paramagnetic limit~\cite{clogston1,maki1,dieterich1}.  Continued
survival of the condensate requires the formation of a 
lower energy spatially-inhomogeneous phase, in 
which pairing is between spin-polarized quasiparticle 
states~\cite{dieterich1,fulde1,larkin1,zanchi1,mckenzie1}. 
The existence of such a phase
in superconductors becomes questionable owing to the
field-induced kinetic energy of orbital currents, which often
suppresses superconductivity more
strongly~\cite{norman1,movshovich1}.  
By contrast, pure CDW systems are free from
orbital currents, yet their high condensation energies increase the
demand on magnetic field strength required to 
reach the Pauli paramagnetic limit~\cite{gruner1}. 
(Per)$_2$Au(mnt)$_2$ is a rare example where this
limit ($\approx 37$~T) falls within reach of the highest available
quasi-static magnetic fields of 45~T~\cite{matos1,graf1}.  In this
paper we use temperatures down to 25~mK (roughly
one-thousandth of the energy gap) to show that
the CDW state surpasses the Pauli paramagnetic
limit, signalling the likely 
appearance of an inhomogeneous phase.

Both superconductivity~\cite{tinkham1} and CDWs~\cite{gruner1} form as
a consequence of electron-phonon interactions.  
Superconductors are
ground states in which gauge symmetry is broken (where the variation
of the magnetic field is dependent on topology)~\cite{tinkham1}, while
CDWs exhibit a periodic charge modulation that breaks translational
symmetry~\cite{gruner1}.  
The BCS (Bardeen-Cooper-Schrieffer)
formalism that applies to superconductivity~\cite{bardeen1}, also
conveniently describes the electronic structure of CDWs, with the gap
in the electronic energy spectrum in the zero temperature limit being
given by $2\Delta_0=\zeta k_{\rm B}T_{\rm c}$, where $T_{\rm c}$ is
the transition temperature.  Whereas the ratio $\zeta=$~3.52 in
weak-coupling BCS theory, $5<\zeta<10$ in CDWs owing to their strong
coupling to the ionic lattice~\cite{gruner1}.  
Upon lowering the
temperature through $T_{\rm c}$, a metal-insulator transition occurs,
below which normal carriers must be thermally excited across the gap
to conduct.  The presence of a 
magnetic field $B$ lowers $T_{\rm c}$~\cite{dieterich1};  
simple theory predicts $T_{\rm c}\rightarrow 0$
at the Pauli paramagnetic limit defined as 
$B_{\rm P}=\Delta_0\sqrt{2}gs\mu_{\rm B}$~\cite{clogston1}, 
where $g$ is the
electron {\it g}-factor, $s$ 
is the electron spin and $\mu_{\rm B}$ is
the electron Bohr magneton.
	
\begin{figure}[htbp]
\centering
\includegraphics[width=8cm]{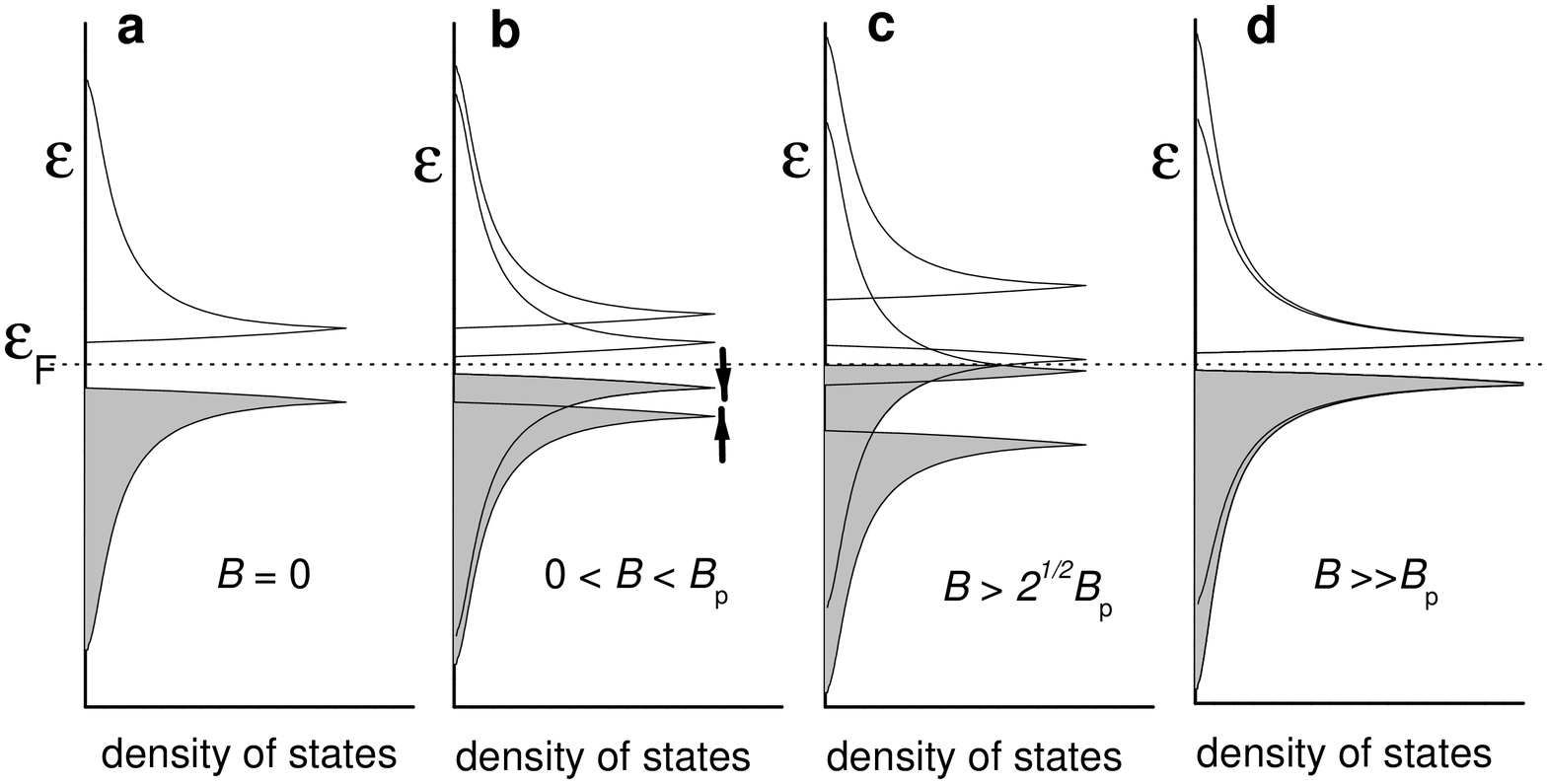}
\caption{The CDW gap in the electronic density of states (DOS) for
different magnetic field strengths $B$.  (a)~The DOS at $B=$~0, with
the shaded region representing occupied states below the Fermi energy
$\varepsilon_{\rm F}$.  (b)~The same DOS for 0~$<B<B_{\rm P}$,
showing closing of the gap.  The Pauli paramagnetic magnetization is
zero since the proportion of spin-up and spin-down states (shown by
arrows) is unchanged from that at $B=$~0.  (c)~The same DOS showing
complete closure of the gap at $B>\sqrt{2}B_{\rm P}$, in which case the
uniform CDW phase cannot be stable.  
(d)~Possible gap formation due to
incommensurate CDW state for $B\gg B_{\rm P}$.  
If the gap stays pinned to
$\varepsilon_{\rm F}$, 
the Pauli paramagnetic magnetization of this state is equivalent
to that of the normal metal.}
\label{gaps}
\end{figure}

Apart from the above energetic considerations~\cite{clogston1}, 
Zeeman splitting
of the electronic bands by $B$ provides another fundamental reason why
uniform CDWs cannot exist for 
$B\gtrsim B_{\rm P}$~\cite{maki1}.  At
$B=0$, 
the optimum modulation vector ${\bf Q}_0$ of the
CDW is equal to the wave vector $2{\bf k}_{\rm F}$ separating states
with opposing momenta $\pm\hbar {\bf k}$ and Fermi velocities $\pm
{\bf v}_{\rm F}$ at the Fermi energy $\varepsilon_{\rm F}$ (the energy
to which the electronic bands are filled)~\cite{gruner1}.  A model of
the density of electronic states gapped upon CDW formation is sketched
in Fig.~\ref{gaps}a.  The Zeeman energy ($\pm gs\mu_{\rm B}B$) causes
$2{\bf k}_{\rm F}$ to differ by for spin-up and spin-down electrons,
causing the spin-up and spin-down energy gaps (formed by ${\bf Q}_0$)
to shift with respect to $\varepsilon_{\rm F}$, as depicted in
Fig.~\ref{gaps}b~\cite{zanchi1,mckenzie1}.  At fields above
$\sqrt{2}B_{\rm P}$ (depicted in Fig.~\ref{gaps}c), $\varepsilon_{\rm
F}$ can no longer reside within the gap~\cite{maki1}.
	
One possible outcome is that the CDW phase is simply destroyed,
reverting to a normal metallic state~\cite{harrison1,graf1}.  Another
more interesting possibility is that the ${\bf Q}_0$ becomes modified
so as to include incommensurate components proportional to the Zeeman
energy so as to create a new gap at $\varepsilon_{\rm F}$, as depicted
in Fig.~\ref{gaps}d~\cite{zanchi1,mckenzie1}. Hence ${\bf Q}_0$ becomes
${\bf Q}_0\pm 2gs\mu_{\rm B}{\bf v}_{\rm F}/\hbar |{\bf v}_{\rm F}|^2$, for
the spin-up and spin-down components respectively, giving rise to
spin-up and spin-down CDWs that are mutually incommensurate which each
other as well as the crystalline lattice.  A simple superposition of
spin-up and spin-down modulations would lead to a combined charge and
spin modulation, with the amplitude further modulated with a very long
period $\lambda=\pi\hbar |{\bf v}_{\rm F}|/2gs\mu_{\rm B}B$, giving rise to
possible nodes.  
While a detailed theoretical model of this complex
inhomogeneous phase has not been made, the modified energy gap
$2\Delta_\lambda$ is expected to be significantly smaller than
$2\Delta_0$~\cite{zanchi1} requiring lower temperatures for its
observation.

Very low temperatures facilitate CDW observation by freezing out
normal carriers that would otherwise be thermally excited across the
energy gap~\cite{gruner1}.  
Electrical conduction can then only take
place via the CDW collective mode, requiring a threshold electric
field $E_{\rm t}$ to depin it from impurities and defects in the
crystalline lattice~\cite{gruner1} (or a threshold voltage $V_{\rm t}$
observed between voltage terminals).  
Once depinned, the CDW is able to slide 
and carry a current with only small
incremental changes in electric field required for large increases in
current.  This gives rise to a distinctive current $I$-versus-voltage
$V$ behaviour that has been observed experimentally in numerous CDW
systems~\cite{gruner1,lopes1}.  The size of the threshold electric
field (or $V_{\rm t}$) depends on the strength of the coupling between
the charge modulation and pinning sites~\cite{gruner1}.  This coupling
is known to become weaker once the CDW becomes more incommensurate or
when the size of the energy gap is reduced.  Both are expected to
occur within the spatially inhomogeneous
phase~\cite{zanchi1,mckenzie1}.
	
The material described in this paper, 
(Per)$_2$Au(mnt)$_2$ belongs to a series of
isostructural charge-transfer salts consisting of one-dimensional
conducting chains of perylene molecules (in the (Per)$_2^+$ oxidation
state) and insulating chains of maleonitriledithiolate (in the
$M$(mnt)$_2^-$ oxidation state), with two formula units per unit cell
giving rise to a 3/4-filled band~\cite{lopes1}.  
CDWs occur for
$M=$~Pt, Cu and Au, with the transition temperature being
approximately 12~K in the $M=$~Au salt.  
This brings $B_{\rm P}$
comfortably within the range of the 45~T Hybrid Magnet at the National
High Magnetic Field Laboratory in Tallahassee~\cite{matos1,graf1}. 
Although a lower transition ( 8~K) occurs in the $M=$~Pt salt, CDW
formation there is compounded by the synchronous formation of a spin
Peierls state involving localized spins on the Pt sites~\cite{matos1}. 
A pure CDW state occurs only in the case of the $M=$~Au and Cu salts.
	
\begin{figure}[htbp]
\centering
\includegraphics[width=8cm]{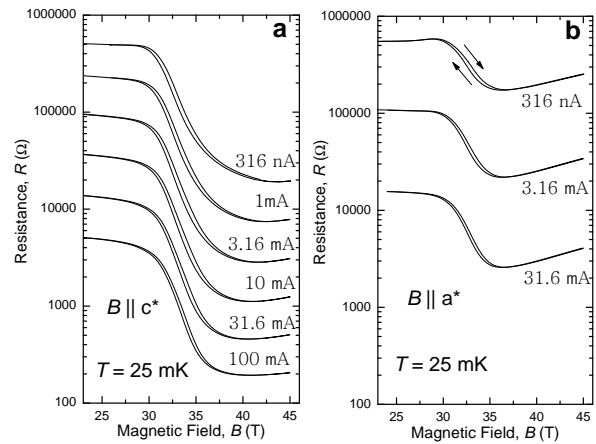}
\caption{Electrical resistance of a single crystal of 
(Per)$_2$Au(mnt)$_2$
measured in a portable dilution refrigerator at 25~mK for
fields between 23 and 45~T, for two different orientations $c^\ast$ (a) and
$a^\ast$ (b) of $B$ perpendicular to its long axis $b$, at several different
applied currents.  The lowest resistance for a given current occurs
for $B$ parallel to $c^\ast$, which is perpendicular to $a^\ast$.  The dependence of
the resistance on current signals non-ohmic behaviour (see Fig.  3). 
Hysteresis between rising and falling magnetic fields (shown by
arrows) is the consequence of a first order Pauli phase transition
between low-magnetic-field uniform CDW and high-magnetic-field
inhomogeneous CDW phases.  Orbital effects involving neighbouring
perylene chains are proposed to account for the field orientation
dependence of the resistance~\cite{graf1}.}
\label{resistance}
\end{figure}

Previous experimental studies of (Per)$_2$Au(mnt)$_2$ have shown the
suppression of the transition 
temperature into the insulating state to
be proportional to the square of the magnetic field ($B^2$) to leading
order, in accordance with the predictions of mean field
theory~\cite{matos1}.  
The CDW energy gap is suppressed in a
qualitatively similar manner with magnetic field, with excitations of
normal carriers across the gap giving rise to a thermally activated
resistance of the form 
$\rho\propto\exp(-\Delta/2k_{\rm B}T)$ on
entering the CDW phase.  
Fits of the resistance to the thermal
activation model enable one to anticipate gap closure of the uniform
CDW phase at $B_{\rm P}\approx$~37~T~\cite{graf1}.  
Assuming
$gs=$~1, this yields $\zeta\approx$~6, 
which falls within the range
typical for CDW ground states~\cite{gruner1}.  
Figure~\ref{resistance} compares plots of the $B$-dependence of the
resistance $R$ along the chains of a needle-shaped sample of
(Per)$_2$Au(mnt)$_2$ (of dimensions
3~mm~$\times$~30~$\mu$m~$\times$~20~$\mu$m), for different values of
the current $I$ and for 
two orthogonal directions of $B$ oriented perpendicular to
the chains at 25~mK. 
The hysteresis between up and down $B$ sweeps
could be the consequence of a first order phase transition, compounded
by CDW pinning effects~\cite{gruner1}.  
The Pauli limit is expected to
yield a first order phase transition~\cite{dieterich1,mckenzie1}.  

\begin{figure}[htbp]
\centering
\includegraphics[width=8cm]{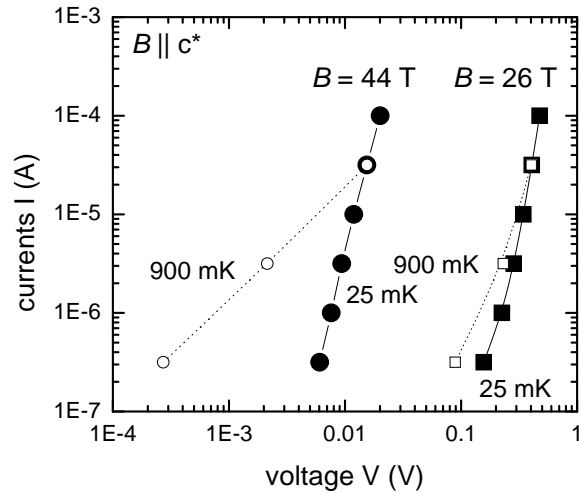}
\caption{Non-linear current-versus-voltage characteristic of
(Per)$_2$Au(mnt)$_2$ plotted on a log-log scale, for selected magnetic
fields (26 and 44~T) above (circles) and below (squares) $B_{\rm P}$. 
Filled symbols connected by solid lines represent data taken at 25~mK
while open symbols connected by dotted lines represent data taken at
900~mK. A strong increase in current for a small increase in voltage
(or electric field) is the characteristic electrodynamic behaviour of
CDWs at very low temperatures where normal carriers are frozen out~\cite{gruner1}. 
The threshold depinning voltage $V_{\rm t}$ is
usually defined as the lowest voltage at which non-linear
$I$-versus-$V$ behaviour is observed, 
yet attempts to drive currents
of 100~nA or smaller through the sample 
were unsuccessful at 25~mK for $B<$~37~T: hence, 
we loosely define $V_{\rm t}\approx$~150~mV at
$B=$~26~T and $\approx$~6~mV at $B=$~44~T
for voltage contacts $\sim$~1~mm apart.}
\label{IV}
\end{figure}

In agreement with
previous studies~\cite{graf1}, the sample resistance $R$ 
shown in Fig.\ref{resistance}, is observed to
drop by roughly an order of magnitude between 28 and 37~T.
This abrupt drop in resistance on surpassing $B_{\rm P}$
was recently interpreted (Ref.\cite{graf1}) as the
destruction of the CDW phase, 
followed by the recovery of
metallic behaviour.  Unfortunately 
the experiments in Ref.~\cite{graf1}
were limited to temperatures
$T\gtrsim$~0.5~K; hence, the high magnetic field region of non-linear
conductivity could not be accessed in those experiments~\cite{graf1}, 
preventing the observation of the true nature of
the ground state at $B>B_{\rm P}$.  
In the present
study, the continued strong dependence of $R$ on $I$ at $B>B_{\rm P}$
at dilution refrigerator temperatures ($T\ll$~0.5~K) in
Fig.~\ref{resistance} is clearly uncharacteristic of a metal.  It is,
nevertheless, quite consistent with the continued presence of a CDW
phase.  The data in Fig.\ref{resistance} display the typical CDW
non-linear $I$-versus-$V$ electrodynamics (where $V=IR$) at all
magnetic fields, characterised by an almost order of magnitude drop in
$R$ for an order of magnitude increase in $I$.  
This becomes
particularly clear on comparing the $I$-versus-$V$ plots at 26 and
44~T in Fig.~\ref{IV}.  The only qualitative change between high and
low magnetic field regimes at 25~mK is that the $I$-versus-$V$ curve
is shifted to lower voltages at magnetic fields above $B_{\rm P}$,
corresponding to a drop in the threshold voltage $V_{\rm t}$ of more
than one decade.  We can therefore state that 
the CDW ground state survives 
$B_{\rm P}$, but with its pinning to the lattice becoming considerably
weakened.  Such weakening is likely to be a consequence of the CDW
becoming increasingly incommensurate on accommodating Zeeman
contributions to the modulation
vector(s)~\cite{dieterich1,zanchi1,mckenzie1}, 
combined with a greatly
reduced energy gap $2\Delta_\lambda$~\cite{zanchi1,mckenzie1}. 

Evidence for the reduced value of $\Delta$ 
is obtained by repeating the $I$-versus-$V$
plot at $T=$~900~mK in Fig.~\ref{IV}.  
This elevated temperature is
approximately 60 times lower than $2\Delta_0$ and is therefore unable
to excite significant numbers of carriers across the gap within the
uniform low magnetic field CDW phase~\cite{gruner1}.  It is, however,
sufficiently high to restore ohmic ($I\propto V$) behaviour
above $B_{\rm P}$.  Ohmic behaviour is restored whenever the gap is
destroyed or when a significant number of carriers are thermally
excited across a gap that has become considerably reduced.  The data
are therefore consistent with mean field theory, which predicts
$2\Delta_\lambda\ll 2\Delta_0$~\cite{zanchi1}.

The survival of CDW electrodynamics to fields far above the Pauli
limit of (Per)$_2$Au(mnt)$_2$, in a region where the uniform phase
cannot exist, indicates the likely development of an inhomogeneous CDW
phase that is stable only at high magnetic fields.  The weakened
threshold electric field for depinning the CDW is consistent with such
a fragile incommensurate phase.  This finding, combined with the
restoration of ohmic behaviour at 900~mK (a temperature that is still
low for the uniform CDW), is consistent with a greatly reduced gap
$2\Delta_\lambda$.  One distinct advantage of the inhomogeneous phase
of a CDW, over that anticipated in superconductors, is that the long
range incommensurate charge and spin modulations can be verified
directly by means of x-ray and neutron diffraction
techniques~\cite{gruner1}.  Until such techniques become available in
fields of $B>$~30~T, however, nuclear-magnetic-resonance may provide
an alternative means for probing incommensurate
structures~\cite{gruner1}.  A direct verification of the inhomogeneous
CDW phase would greatly enhance our understanding of the generic
properties of singlet-paired condensates in high magnetic fields.  It
would also provide an essential precedent for understanding proposed
inhomogeneous phases in $s$- and $d$-wave paired
superconductors~\cite{fulde1,larkin1,norman1,movshovich1}.

This work was performed under the auspices of the National Science
Foundation, Department of Energy and the State of Florida.  We would
like to thank Timothy Murphy and Eric Palm for their help with the
portable dilution refrigerator.


\begin{thebibliography}{99}

\bibitem{malo1} C.~A.~R.~S\'{a}~de~Malo (ed.), {\it The
superconducting state in magnetic fields}
(World Scientific, Singapore, 1998).%1

\bibitem{chaikin1} P.~M.~Chaikin, J. Phys.  I (France) {\bf 6}, 1875
(1996).%2

\bibitem{singleton1} J.~Singleton, Rep.  Prog.  Phys.  {\bf 63}, 1111
(2000).%3

\bibitem{gruner1} G.~Gr\"{u}ner, {\it Density waves in solids, Frontiers 
in physics {\bf 89}} (Addison-Wesley, 1994).%4

\bibitem{tinkham1} M.~Tinkham {\it Introduction to superconductivity, Second Edition}
(McGraw-Hill, 1996).%5

\bibitem{clogston1} A.~M.~Clogston, Phys. Rev. Lett. {\bf 9}, 266 
(1962).%6

\bibitem{maki1} K.~Maki and T.~Tsuneto, Prog. Theor. Phys. {\bf 31}, 945 
(1964). %7

\bibitem{dieterich1} W.~Dieterich and P.~Fulde, Z. Phys. {\bf 265}, 238 
(1973).%8

\bibitem{harrison1} N.~Harrison, Phys. Rev. Lett. {\bf 83}, 1395 
(1999).%9

\bibitem{fulde1} P.~Fulde and R.~A.~Ferrel, Phys. Rev. {\bf 135}, A550 
(1964).%10

\bibitem{larkin1} A.~I.~Larkin and Y.~N.~Ovchinnikov, Sov. Phys. JETP 
{\bf 20}, 762 (1965).%11

\bibitem{zanchi1} D.~Zanchi, A.~Bjelis and G.~Montambaux, Phys. Rev. B 
{\bf 53}, 1240 (1996).%12

\bibitem{mckenzie1} R.~H.~McKenzie, (unpublished) cond-mat/9706235.%13

\bibitem{norman1} M.~R.~Norman, Phys. Rev. Lett. {\bf 71}, 3391 
(1993).%14

\bibitem{movshovich1} R.~Movshovich, A.~Bianchi, C.~Capan, M.~Jaime and 
R.~Goodrich, (comment submitted to Nature, 2003).%15

\bibitem{matos1} M.~Matos, G.~Bonfait, R.~T.~Henriques, and M.~Almeida, 
Phys. Rev. B {\bf 54}, 15307 (1996).%16

\bibitem{graf1} D.~Graf, J.~S.~Brooks, E.~S.~Choi, J.~C.~Dias, M.~Almeida, 
and M.~Matos, (unpublished) cond-mat/0311399.%17

\bibitem{bardeen1} J.~Bardeen, L.~N.~Cooper, and J.~R.~Schrieffer, Phys. 
Rev. {\bf 108}, 1175 (1957).%18

\bibitem{lopes1} E.~B.~Lopes, M.~J.~Matos, R.~T.~Henriques, M.~Almeida, 
and J.~Dumas, J. Phys. I France {\bf 6}, 2141 (1996).%19



\end{thebibliography}
\end{document}